\documentclass[twocolumn,showpacs,preprintnumbers,amsmath,amssymb]{revtex4}
\usepackage{graphicx}
\usepackage{dcolumn}
\usepackage{bm}
\begin{document}
\title{Unconventional motional narrowing in the optical spectrum of a semiconductor quantum dot}
\author{A. Berthelot$^1$, I. Favero$^1$, G. Cassabois$^{1,\ast}$, C. Voisin$^1$, C.
Delalande$^1$, Ph. Roussignol$^1$, R. Ferreira$^1$ and J. M.
G\'{e}rard$^2$} \affiliation{$^1$Laboratoire Pierre Aigrain, Ecole
Normale
Sup\'erieure, 24 rue Lhomond 75231 Paris Cedex 5, France\\
$^2$CEA-CNRS-UJF "Nanophysics and Semiconductors" Laboratory,
CEA/DRFMC/SP2M, 17 rue des Martyrs 38054 Grenoble Cedex 9, France}
\date{\today}
\begin{abstract}
Motional narrowing refers to the striking phenomenon where the
resonance line of a system coupled to a reservoir becomes narrower
when increasing the reservoir fluctuation. A textbook example is
found in nuclear magnetic resonance, where the fluctuating local
magnetic fields created by randomly oriented nuclear spins are
averaged when the motion of the nuclei is thermally activated. The
existence of a motional narrowing effect in the optical response
of semiconductor quantum dots remains so far unexplored. This
effect may be important in this instance since the decoherence
dynamics is a central issue for the implementation of quantum
information processing based on quantum dots. Here we report on
the experimental evidence of motional narrowing in the optical
spectrum of a semiconductor quantum dot broadened by the spectral
diffusion phenomenon. Surprisingly, motional narrowing is achieved
when decreasing incident power or temperature, in contrast with
the standard phenomenology observed for nuclear magnetic
resonance.
\end{abstract}

\maketitle

In the seminal work on motional narrowing by Bloembergen
\textit{et al.}, relaxation effects in nuclear magnetic resonance
were beautifully explained by taking into account the influence of
the thermal motion of the magnetic nuclei upon the spin-spin
interaction \cite{bloembergen}. The general treatment of
relaxation processes for a system interacting with a reservoir was
later formulated by Kubo in a stochastic theory that assumes
random perturbations of the system by a fluctuating environment
\cite{kubo}. Depending on the relative magnitude of the spectral
modulation amplitude and the inverse of the modulation correlation
time, the relaxation dynamics is either in the \textit{slow
modulation} limit, where the optical line-shape reflects directly
the statistical distribution of the different system energies, or
in the \textit{fast modulation} limit where the fluctuation is
smoothed out and the line-shape is motionally narrowed into a
Lorentzian profile. The relevance of motional narrowing for the
description of relaxation phenomena has spread throughout many
different fields, such as spin relaxation in semiconductors
\cite{dyakonov}, vibrational dephasing in molecular physics
\cite{oxtoby}, or phase noise in optical pumping \cite{eberly}.

The optical spectrum of a material system with localized,
zero-dimensional electronic states provides a generic example of
the influence of a fluctuating environment on the coherence
relaxation dynamics. In that case, the perturbing interactions
induce a stochastic shift over time of the optical spectrum,
resulting in the so-called spectral diffusion effect, which was
observed for rare-earth ions \cite{flach}, molecules
\cite{moerner}, or semiconductor quantum dots
\cite{bawendi,robinson}. In this latter system, impurities,
defects or localized charges in the vicinity of a quantum dot
induce micro-electric fields that shift the quantum dot emission
line through the quantum confined Stark effect. The fluctuation of
the quantum dot environment thus randomize the emission energy
over a spectral range $\Sigma$ on a characteristic time scale
$\tau_c$. Spectral diffusion under light illumination was reported
with jitters of the quantum dot emission energy from hundreds
$\mu$eV to few meV on time scales ranging from milliseconds to
minutes \cite{bawendi,robinson,besombes}. However, to the best of
our knowledge, no evidence for motional narrowing
($\Sigma\tau_c$$\ll$$\hbar$) was pointed out for the spectral
diffusion phenomenon in quantum dots.

We present here the experimental evidence for motional narrowing
in the optical spectrum of a semiconductor quantum dot.
High-resolution Fourier-transform spectroscopy performed on the
photoluminescence signal of a single InAs/GaAs quantum dot allows
the determination of both width and shape of the emission line. A
crossover from Lorentzian to Gaussian profiles of the zero-phonon
line is observed while its linewidth increases. We obtain a
quantitative agreement with Monte Carlo simulations where the
fluctuating environment is modelled by the random capture and
escape of carriers in traps located in the quantum dot vicinity.
We show that motional narrowing is achieved when decreasing the
temperature or the incident power because of the asymmetry of the
capture and escape mechanisms. Our study provides a
counterintuitive example of motional narrowing compared to the
standard phenomenology described in nuclear magnetic resonance.

We study self-assembled InAs/GaAs quantum dots grown by molecular
beam epitaxy in the Stranski-Krastanow mode, and
micro-photoluminescence measurements under non-resonant excitation
are performed in the far field using the experimental setup
described in Ref.~\cite{kammerer}. In order to accurately
characterize the quantum dot emission spectrum, the
photoluminescence signal arising from a single quantum dot is
analyzed by high-resolution Fourier-transform spectroscopy
\cite{kammerer}. The Fourier-transform technique is implemented in
the detection part of the setup where the photoluminescence signal
passes through a Michelson interferometer placed in front of a 32
cm grating spectrometer. The signal is detected by a low noise
Si-based photon counting module. With a translation stage, we vary
the time $t$ for propagation in one arm of the interferometer and
record interferograms of the photoluminescence emission
$I(t)$=$I_0$$(1+C(t)\cos(E_{0}t/\hbar))$, where $I_0$ is the
average photoluminescence signal intensity, $E_0$ the central
detection energy, and $C(t)$ the interference contrast. In the two
cases of Lorentzian or Gaussian line-shapes, a Fourier-transform
analysis shows that the interference contrast decay is exponential
or Gaussian, respectively. As we will see below, the
implementation of this technique in single quantum dot
spectroscopy allows an accurate determination of both width and
shape of the emission line. As far as the linewidth is concerned,
sub-$\mu$eV measurements can be performed by using this
interferometric correlation technique \cite{kammerer}. For the
determination of the line-shape, our technique allows a
high-resolution sampling of the Fourier-transform of the spectrum
on typically thousands of points. This value is by two orders of
magnitude larger than the average number of illuminated pixels in
a charge-coupled device in the case of a multi-channel detection
in the spectral domain. The drawbacks of our Fourier-transform
technique are the long acquisition times of roughly one hour per
interferogram, and the required extremely high mechanical
stability of the system during the interferogram acquisition.
\begin{center}
\begin{figure}[h,t]
\includegraphics[scale=0.8,angle=0]{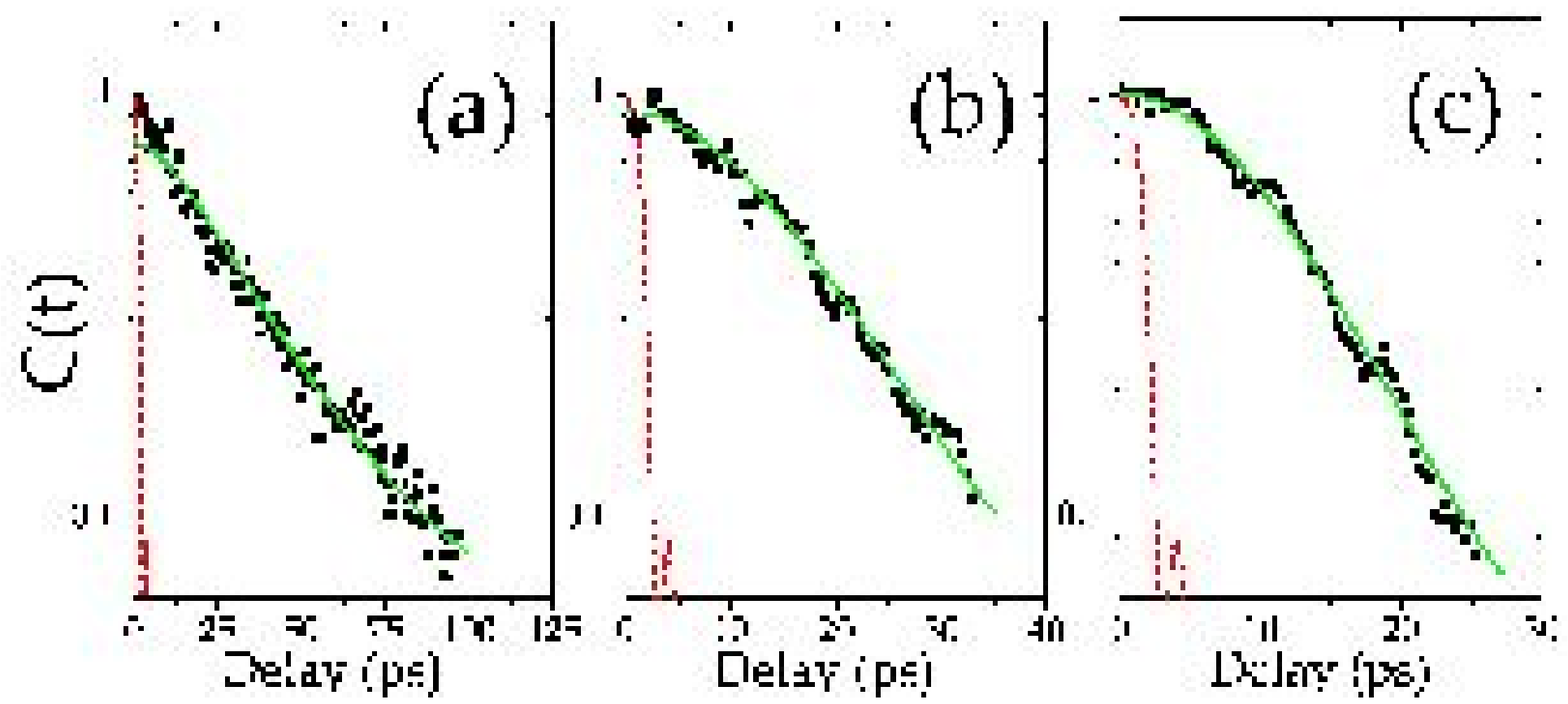}
\caption{Interferogram contrast $C(t)$ of the photoluminescence
signal of a single InAs/GaAs quantum dot at 10K, on
semi-logarithmic plots, for three different incident powers: 0.18
(a), 0.72 (b), and 2.88 kW.cm$^{-2}$ (c). Data (squares), system
response function (dotted line, in red), theoretical fits (solid
line, in green) obtained by the convolution of the system response
function with Eq.~(\ref{eq1}), with $\Sigma\tau_c$/$\hbar$$\sim$
0.6 (a), 1.05 (b), and 1.35 (c), are plotted as a function of the
delay $t$.} \label{fig1}
\end{figure}
\end{center}

In Fig.~\ref{fig1} we display the measured (squares) interference
contrast $C(t)$ for the emission spectrum of a single InAs/GaAs
quantum dot at 10K, on semi-logarithmic plots, for three different
incident powers: 0.18 (a), 0.72 (b), and 2.88 kW.cm$^{-2}$ (c). We
first observe that the coherence relaxation dynamics becomes
faster when increasing the incident power. Moreover, we notice a
gradual modification in the shape of $C(t)$. At low power
(Fig.~\ref{fig1}(a)), the interference contrast decay is
quasi-exponential with a time constant of 29 ps, thus
corresponding to a quasi-Lorentzian profile with a full width at
half maximum (FWHM) of 45 $\mu$eV. At higher power
(Fig.~\ref{fig1}(b)), $C(t)$ has a Gaussian decay at short times
($t$$\leq$15 ps), and remains exponential at longer times. The
line has an intermediate profile, and here its FWHM is 105
$\mu$eV. At the highest incident power (Fig.~\ref{fig1}(c)), the
interference contrast decay is predominantly Gaussian, thus
corresponding to a quasi-Gaussian profile, with a FWHM of 155
$\mu$eV.

This crossover from a Lorentzian to a Gaussian line-shape when
increasing the incident power reveals an original relaxation
dynamics in semiconductor physics. In fact, in bulk semiconductors
or in quantum wells, the standard phenomenology is the exact
opposite. An increasing number of photo-created carriers leads to
the activation of carrier-carrier Coulomb correlations, and
induces the inverse transition from an inhomogeneously broadened
Gaussian line to a homogeneous Lorentzian one \cite{shah}. In the
following, we provide a quantitative analysis of our measurements
in the framework of the Kubo theory.

Applying the stochastic theory of line-shape and relaxation of a
system coupled to a fluctuating reservoir, the emission spectrum
of a quantum dot is conveniently characterized in the time domain
and, assuming Gaussian fluctuation, the Fourier-transform of the
intensity spectrum has the general expression \cite{kubo}:
\begin{equation}
C(t)=\exp\left[-\frac{\Sigma^2\tau_c
^2}{\hbar^2}\left(\exp\left(-\frac{t}{\tau_c}\right)+\frac{t}{\tau_c}-1\right)\right]
\label{eq1}
\end{equation}
In the limit of \textit{slow modulation}
($\Sigma\tau_c$$\gg$$\hbar$), the characteristic function $C(t)$
has a Gaussian decay ($C(t)$$\sim$$\exp(-\Sigma^2t^2/2\hbar^2)$):
the spectrum reflects directly the statistical distribution of the
emission energies, and the line-shape is Gaussian with a FWHM
given by $2\sqrt{2ln2}$\,$\Sigma$. In the opposite limit of
\textit{fast modulation} ($\Sigma\tau_c$$\ll$$\hbar$), the
motional narrowing effect gives rise to an exponential decoherence
($C(t)$$\sim$$\exp(-\Sigma^2\tau_c t/\hbar^2)$) with a decay time
$T_2$=$\hbar^2/\Sigma^2\tau_c$ : the fluctuation is smoothed out
by the fast modifications of the reservoir configuration, and the
line-shape becomes Lorentzian with a FWHM given by
$2\Sigma^2\tau_c/\hbar$. Indeed, when motional narrowing occurs,
the linewidth is reduced by a factor of the order of
$\Sigma\tau_c$/$\hbar$.

A comparison between our experimental data and the general
expression of $C(t)$ given above allows us to characterize the
transition from the exponential to the Gaussian relaxation
observed in Fig.~\ref{fig1}. For this purpose, we convolute
Eq.~(\ref{eq1}) with the system response function (dotted lines in
Fig.~\ref{fig1}) which is obtained under white light illumination.
The calculated fits are displayed in solid line in
Fig.~\ref{fig1}. We obtain an excellent agreement with increasing
values of $\Sigma\tau_c$/$\hbar$ of 0.6 in (a), 1.05 in (b), and
1.35 in (c), where the correlation time $\tau_c$ has a constant
value of 10 ps. The gradual increase of $\Sigma\tau_c$/$\hbar$
demonstrates the transition from the \textit{fast modulation}
limit to the \textit{slow modulation} one when increasing the
incident power. As a matter of fact, it shows that motional
narrowing occurs for low excitation. At first glance, one may
think that increasing the number of photo-created carriers in the
structure should induce some additional motion in the environment,
so that the quantum dot line should be motionally narrowed in the
high excitation limit. It is in fact the exact opposite, and we
explain below the origin of this unconventional phenomenology.
\begin{center}
\begin{figure}[h,t]
\includegraphics[scale=0.95,angle=0]{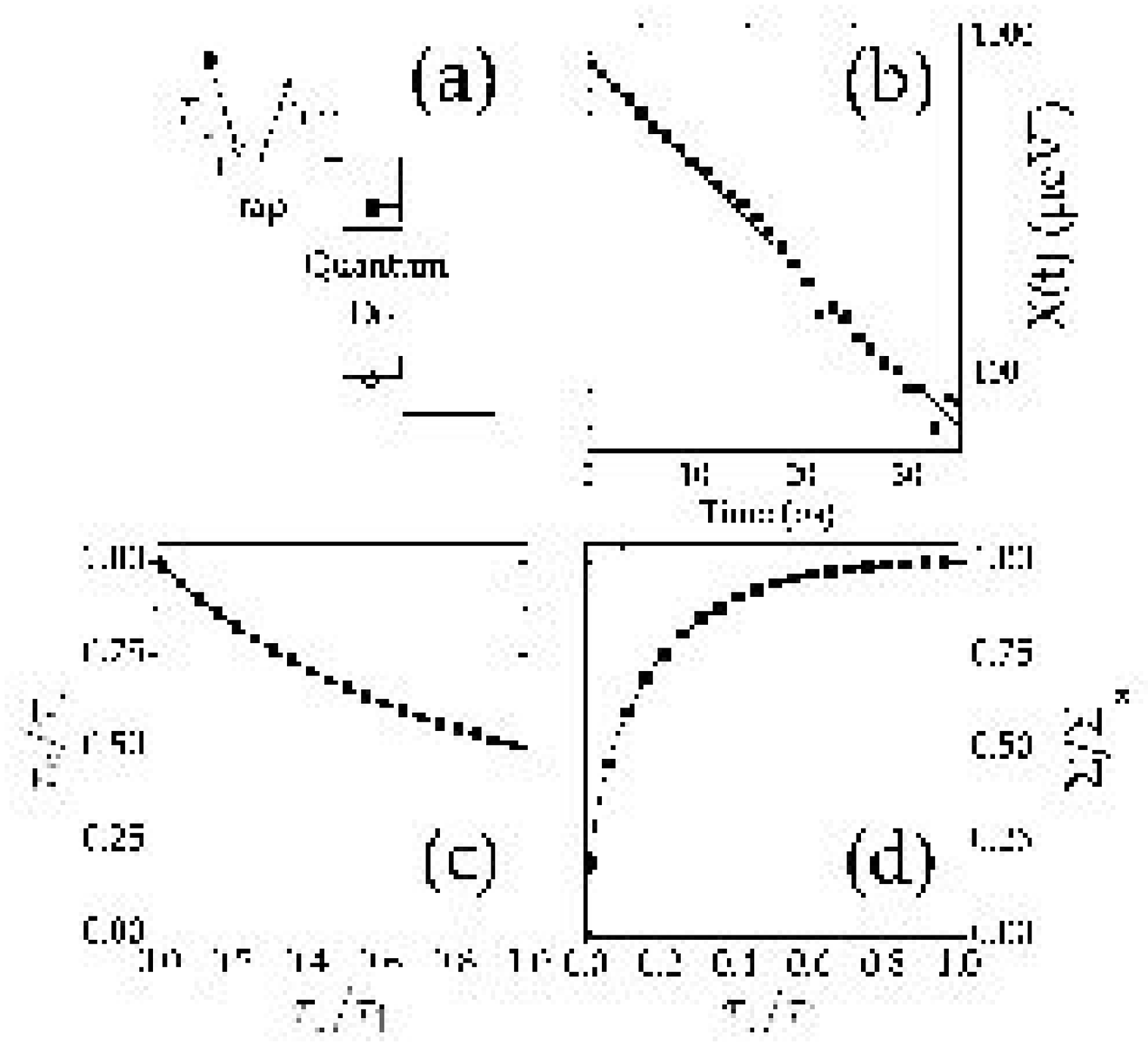}
\caption{(a) Schematic of the quantum dot and its fluctuating
environment: $\tau_\downarrow$ and $\tau_\uparrow$ are the capture
and escape times of carriers. (b) Correlation function of the
energy fluctuation as a function of time, on a semi-logarithmic
plot: Monte-Carlo simulations (squares) and exponential fit (solid
line). (c) Correlation time $\tau_c$ normalized to the capture
time $\tau_\downarrow$ versus $\tau_\downarrow$/$\tau_\uparrow$
and (d) spectral modulation amplitude $\Sigma$ normalized to its
saturation value $\Sigma_c$ versus
$\tau_\downarrow$/$\tau_\uparrow$: Monte-Carlo simulations
(squares), and fits (solid lines) according to Eq.~(\ref{eq2}) for
(c) and Eq.~(\ref{eq3}) for (d).} \label{fig2}
\end{figure}
\end{center}

The existence of a fluctuating environment around the quantum dot
originates from the presence of impurities, defects in the barrier
material or in the wetting layer. Furthermore, this latter
heterostructure is far from being an ideal two-dimensional quantum
well so that the unavoidable interface roughness gives also rise
to localized states in the quantum dot surroundings
\cite{robinson,gerard,heitz}. The characteristic parameters
$\Sigma$ and $\tau_c$ are thus determined by the population
fluctuation in the various traps located around the quantum dot.
We present Monte Carlo simulations of the carrier dynamics in the
reservoir for the understanding of the motional narrowing effect
in quantum dots.

In our model, the fluctuating reservoir consists in $N$
uncorrelated traps which individually induce a Stark shift
$\Delta$ of the quantum dot line. This simple scheme has been
chosen for the sake of clarity. More realistic geometries can be
used but do not alter the conclusions. Each trap is loaded with a
characteristic capture time $\tau_\downarrow$, and emptied with an
escape time $\tau_\uparrow$ (Fig.~\ref{fig2}(a)). In our
time-dependent simulations, the random modifications of the traps
population result in fluctuation over time $\delta E(t)$ of the
transition energy around a mean value $E_0$. Then, the spectral
modulation amplitude $\Sigma$ and the correlation time $\tau_c$
are calculated by computing the correlation function of the energy
fluctuation $X(t)$, which is defined by the configuration average
$X(t)$=$\langle\delta E(t)\delta E(0)\rangle$ \cite{kubo}. This
correlation function is expected to have an exponential decay
given by $X(t)$=$\Sigma^2$$\exp(-t/\tau_c)$ \cite{kubo}. In
Fig.~\ref{fig2}(b), we display the correlation function $X(t)$
computed for $N$=50 and averaged over one thousand configurations.
We observe on this semi-logarithmic plot the exponential decay of
$X(t)$, from which we extract $\Sigma$ and $\tau_c$. In
Fig.~\ref{fig2}(c) we display the correlation time $\tau_c$
normalized to the capture time $\tau_\downarrow$ as a function of
the ratio $\tau_\downarrow$/$\tau_\uparrow$. We observe that our
calculations are fitted by using the simple expression:
\begin{eqnarray}
\frac{1}{\tau_c}&=&\frac{1}{\tau_\uparrow}+\frac{1}{\tau_\downarrow}
\label{eq2}
\end{eqnarray}
which shows that the correlation time is inversely proportional to
the total fluctuation rate of the traps population.

Finally, in Fig.~\ref{fig2}(d), we display the spectral modulation
amplitude $\Sigma$ normalized to its saturation value $\Sigma_s$.
Our numerical simulations are well reproduced by using the
expression:
\begin{eqnarray}
\Sigma&=&\frac{\sqrt{N}\Delta}{\sqrt{\frac{\tau_\uparrow}{\tau_\downarrow}}+\sqrt{\frac{\tau_\downarrow}{\tau_\uparrow}}}
\label{eq3}
\end{eqnarray}
which can also be analytically derived by assuming ergodicity of
the reservoir. The saturation value $\Sigma_s$ of the spectral
modulation amplitude is thus given by $\sqrt{N}\Delta/2$. In
semiconductor nanostructures, at low temperature carrier escape is
known to be much less efficient than the capture process
\cite{shah}. When $\tau_\downarrow$/$\tau_\uparrow$$\ll$1, the
correlation time $\tau_c$ has approximately a constant value given
by $\tau_\downarrow$ whereas $\Sigma$ scales as
2$\Sigma_s$$\sqrt{\tau_\downarrow/\tau_\uparrow}$. Therefore, an
activation of the escape process by raising the incident power
will lead to an \textit{increase} of $\Sigma\tau_c$/$\hbar$, in
qualitative agreement with the phenomenology discussed above.

\begin{center}
\begin{figure}[h,t]
\includegraphics[scale=0.8,angle=0]{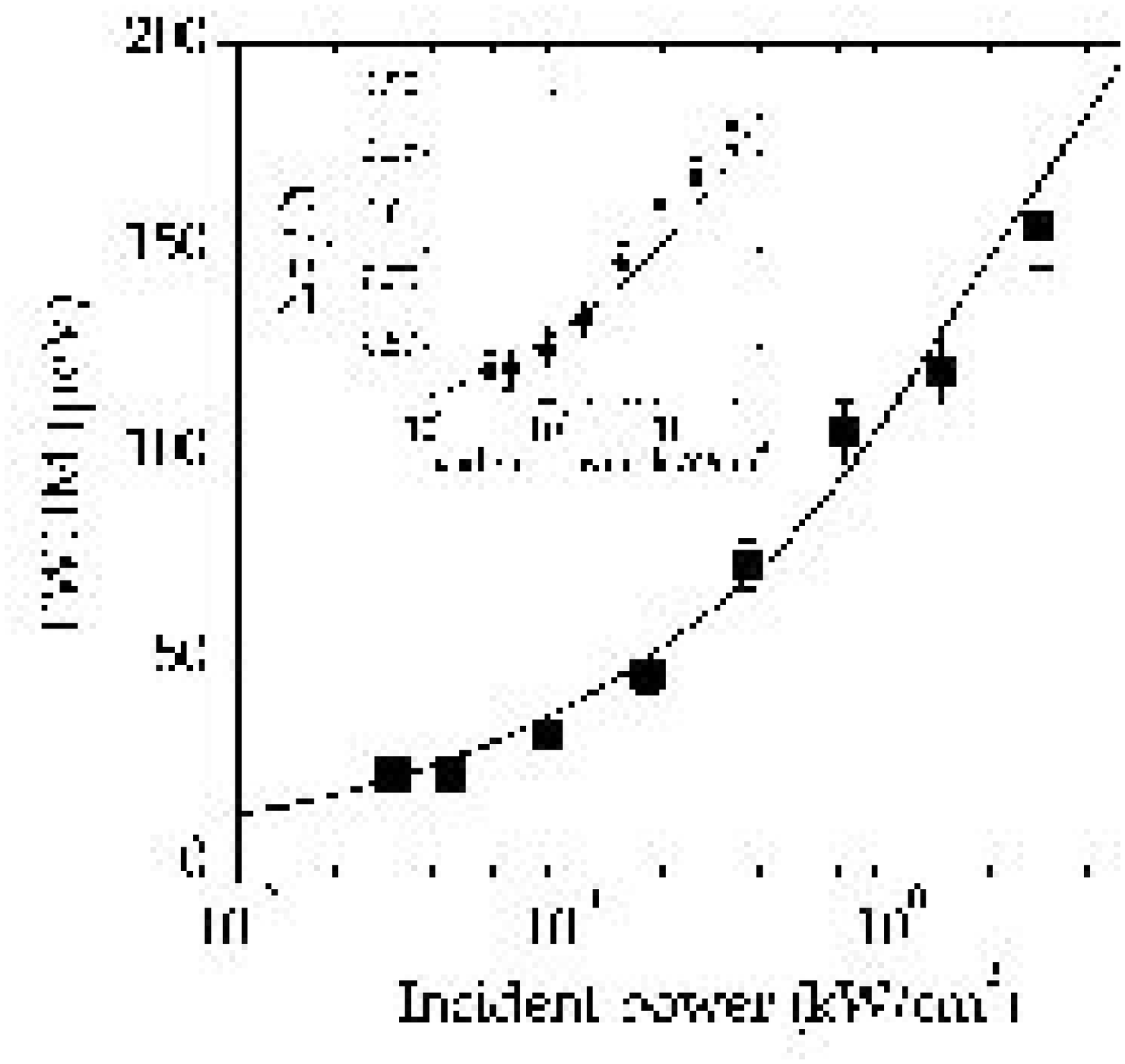}
\caption{Full width at half maximum (FWHM) and
$\Sigma\tau_c$/$\hbar$ (inset) of the zero-phonon line of a single
quantum dot, at 10K. Data (symbols), and calculations (solid
lines) based on Eq.~(\ref{eq1}-\ref{eq3}) are plotted as a
function of incident power, on a semi-logarithmic scale.}
\label{fig3}
\end{figure}
\end{center}

On the basis of our simple microscopic model, we present a
quantitative interpretation of our experiments at 10K as a
function of incident power. In Fig.~\ref{fig3} we display the
measured values of the FWHM (squares) and $\Sigma\tau_c$/$\hbar$
(circles, inset) as a function of the excitation density, on a
semi-logarithmic scale. These two parameters characterize the
width of the line-profile and its shape since the crossover from
the exponential to the Gaussian decoherence dynamics occurs around
$\Sigma\tau_c$/$\hbar$$\sim$1. We confront our data with the
theoretical values (solid lines) of the FWHM and
$\Sigma\tau_c$/$\hbar$ that are calculated with
Eq.~(\ref{eq1}-\ref{eq3}), and we observe a fair agreement for the
set of parameters $\Sigma_s$$\sim$400 $\mu$eV,
$\tau_\downarrow$$\sim$10 ps, and
1/$\tau_\uparrow$=(1/$\tau_0$)$\sqrt{P}$ where $\tau_0$$\sim$1.6
ns and $P$ is in unit of 1 kW.cm$^{-2}$.

The asymmetry between the power dependences of $\tau_\downarrow$
and $\tau_\uparrow$ stems from the existence of different
microscopic processes of various efficiencies in semiconductor
physics \cite{shah}. The constant value of 10 ps for
$\tau_\downarrow$ is consistent with an optical-phonon assisted
capture in the traps around the quantum dot \cite{kaiser}. On the
other hand, the inverse process of carrier escape by
optical-phonon absorption is strongly inhibited at low
temperature, so that it does not significantly contribute to the
value of $\tau_\uparrow$. In fact, the escape rate dependence on
incident power is characteristic of Auger-type processes. In the
elastic collision of two carriers where one is ejected from the
trap while a delocalized carrier relaxes in energy, the escape
rate is proportional to the density of delocalized carriers, which
increases with incident power. In the expression
1/$\tau_\uparrow$=(1/$\tau_0$)$\sqrt{P}$, the sublinear increase
of the escape rate with $P$ can come from the fact that the actual
density of delocalized carriers is governed by bimolecular
interband radiative recombination \cite{bimolecular} or Auger
scattering \cite{auger}, which both result in a square root
dependence of the carrier density on $P$. Lastly, we note that in
the limit of low incident power, the saturation value $\Sigma_s$
given by $\sqrt{N}\Delta/2$ may be power-dependent due to the
reduced number of active traps that are populated from the
reservoir of delocalized carriers. Systematic measurements as a
function of temperature and incident power on several quantum dots
(to be described elsewhere) allow us to discriminate between the
variations of $N$ and $\tau_\uparrow$ in our experiments. The
power-dependence of $N$ is found to be negligible in the
investigated range of incident power, in agreement with the
assumptions made for our calculations.
\begin{center}
\begin{figure}[h,t]
\includegraphics[scale=0.655,angle=0]{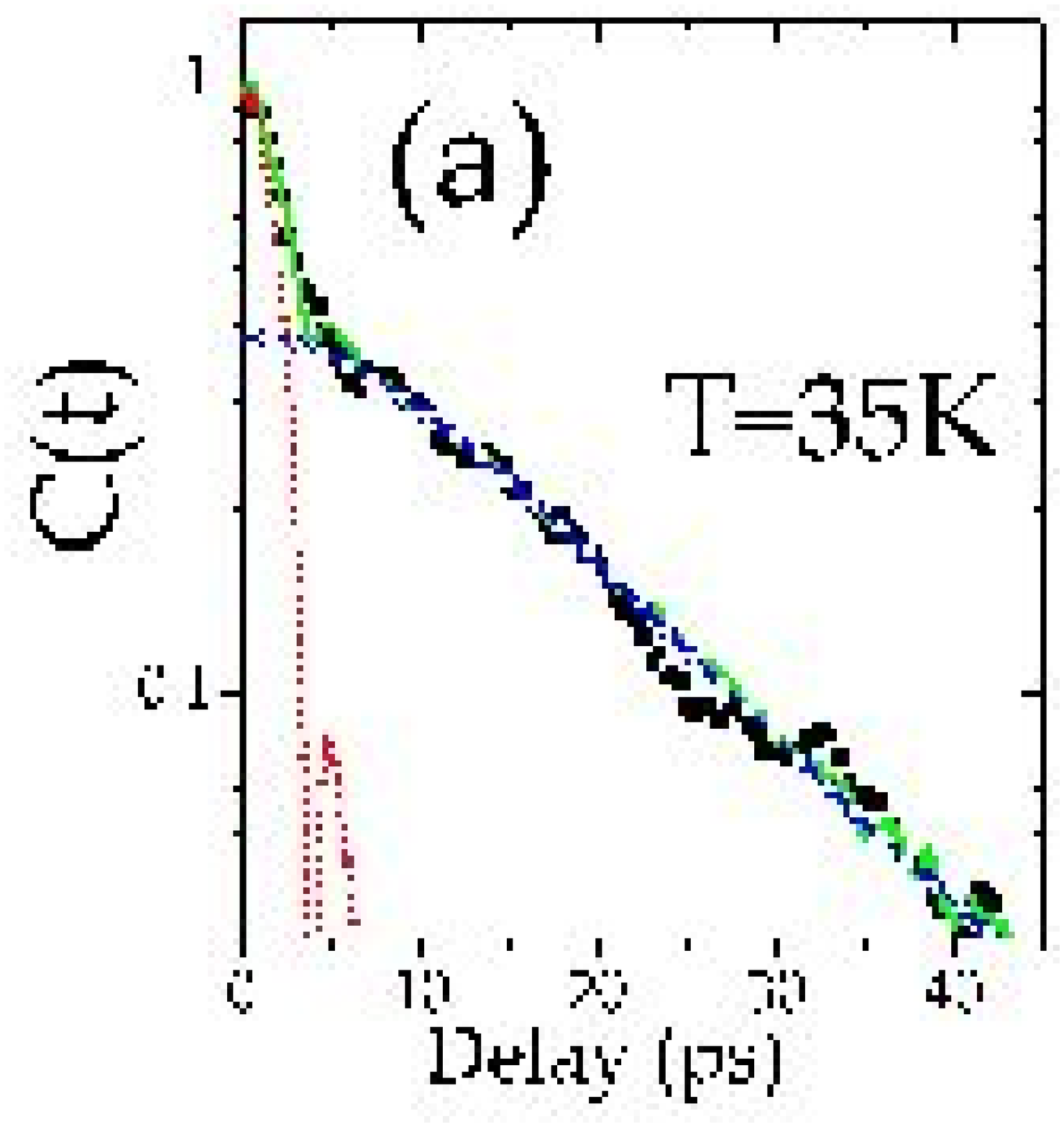}
\includegraphics[scale=0.67,angle=0]{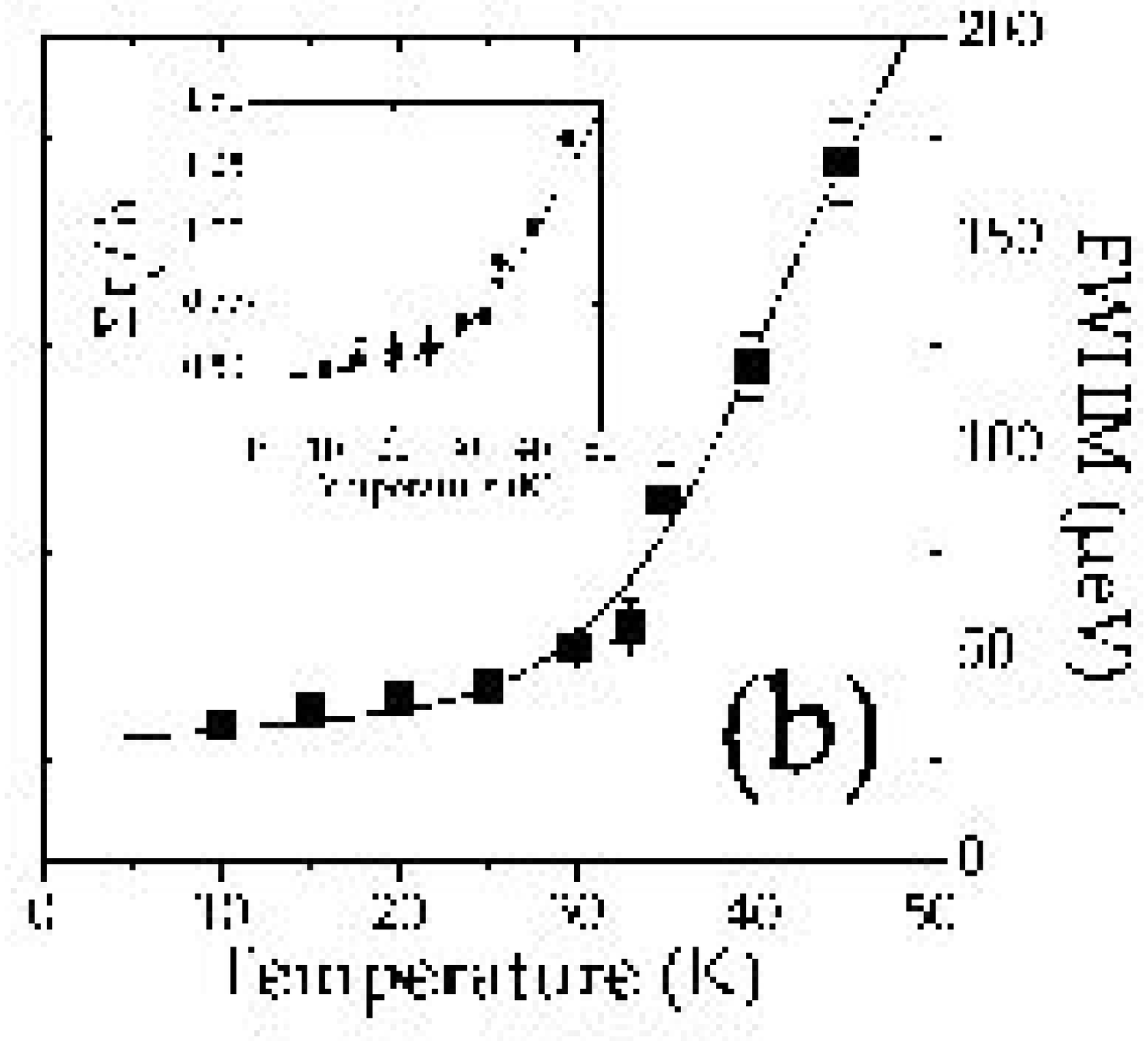}
\caption{(a) Interferogram contrast $C(t)$ of the
photoluminescence signal of the quantum dot of Fig.~\ref{fig3} at
35K, for an incident power of 0.07 kW.cm$^{-2}$. Data (squares),
system response function (dotted line, in red), theoretical fit
(solid line, in green) and zero-phonon line contribution
(line+empty circles, in blue) are plotted as a function of the
delay $t$. The theoretical fit is the convolution of the system
response function with the sum of Eq.~(\ref{eq1}) with
$\Sigma\tau_c$/$\hbar$$\sim$ 0.91 (accounting for the zero-phonon
line contribution) and a $\delta(t)$-function (for the side-bands
background). (b) Full width at half maximum (FWHM) and
$\Sigma\tau_c$/$\hbar$ (inset) of the zero-phonon line of the
quantum dot of Fig.~\ref{fig3}, at 0.07 kW.cm$^{-2}$. Data
(symbols), and calculations (solid lines) based on
Eq.~(\ref{eq1}-\ref{eq3}) are plotted as a function of
temperature.} \label{fig4}
\end{figure}
\end{center}

Temperature-dependent experiments performed at low incident power
(0.07 kW.cm$^{-2}$) reveal the same striking phenomenology, namely
we observe a gradual increase of $\Sigma\tau_c$/$\hbar$ with
temperature which demonstrates that motional narrowing occurs at
low temperature. However, the analysis of the data recorded as a
function of temperature requires a more careful procedure because
the line-profile exhibits broad side-bands around the central
so-called zero-phonon line \cite{besombes2,favero}. These
side-bands are related to the radiative recombination assisted by
the emission and the absorption of acoustic phonons. At low
temperature (T$<$25K), their contribution to the total integrated
intensity of the photoluminescence signal is negligible and can be
safely neglected in the data interpretation, as done for our first
set of measurements as a function of incident power at 10K. Above
25K, the presence of broad lateral side-bands around the
zero-phonon line results in an interference contrast $C(t)$ with
an additional fast component, the decay of which follows our
system response function (Fig.~\ref{fig4}(a)). By adding a
constant term to the line-profile extracted from Eq.~(\ref{eq1}),
we are able to fit the whole decay dynamics of the interference
contrast (solid line, in green in (Fig.~\ref{fig4}(a)), and thus
separate the relative contributions of the lateral side-bands and
zero-phonon line (line+empty circles, in blue in
(Fig.~\ref{fig4}(a)).

In Fig.~\ref{fig4}(b) we display the measured values of the FWHM
(squares) and $\Sigma\tau_c$/$\hbar$ (circles, inset) of the
zero-phonon line as a function of the temperature. In the
temperature-dependent measurements, the crossover from a
Lorentzian to a Gaussian shape of the zero-phonon line occurs when
increasing the temperature. More precisely, a smooth transition
analog to the one presented in Fig.~\ref{fig1} is recorded in the
temperature range between 30K and 45K. We then confront our data
with the theoretical values (solid lines) of the FWHM and
$\Sigma\tau_c$/$\hbar$ that are calculated with
Eq.~(\ref{eq1}-\ref{eq3}), by assuming that:
\begin{eqnarray}
\frac{1}{\tau_\downarrow}&=&\frac{1}{\tau_1}(1+n_1(T))+\frac{1}{\tau_2}(1+n_2(T))\\
\frac{1}{\tau_\uparrow}&=&\frac{1}{\tau_1}n_1(T)+\frac{1}{\tau_2}n_2(T)+\frac{1}{\tau_3}
\label{eq5}
\end{eqnarray}
In Eq.~(\ref{eq5}) the constant term inversely proportional to
$\tau_3$ accounts for the contribution of Auger processes to
carrier escape, as discussed above, and its magnitude is fixed by
the incident power. In the interpretation of our
temperature-dependent measurements, the relevant microscopic
processes for the carrier dynamics in the traps are
phonon-assisted mechanisms where the capture (escape) is
thermally-activated due to the emission (absorption) of an
acoustic phonon of mean energy $E_1$ or an optical phonon of
energy $E_2$. The capture and escape efficiencies are proportional
to ($1$+$n_i(t)$) and $n_i(t)$ respectively, where $n_i(t)$ is a
Bose-Einstein occupation factor given by 1/($\exp(E_i/kT)$-1). In
Fig.~\ref{fig4}(b), we observe a fair agreement between the data
and the calculations for the set of parameters $\Sigma_s$$\sim$400
$\mu$eV, $\tau_1$$\sim$35 ns, $\tau_2$$\sim$10 ps,
$\tau_3$$\sim$6.5 ns, $E_1$$\sim$1 meV, and $E_2$$\sim$20 meV.
This latter activation energy is smaller than the characteristic
optical phonon energy of 36 meV in bulk GaAs, and this difference
may come from strain and confinements effects in the traps formed
in the GaAs barrier or in the InAs wetting layer \cite{jusserand}.
As far as the carrier dynamics in the traps is concerned, we find
that in our experiments performed below 50K, the capture process
is dominated by optical phonon emission
($\tau_\downarrow$$\sim$$\tau_2$). On the other hand, the thermal
activation of the escape one is predominantly due to acoustic
phonon absorption below 40K, with an onset of optical phonon
absorption around 40K. Finally, we conclude that in the
investigated range of temperature, we are always in the regime
where the capture process is much more efficient than the escape
one ($\tau_\downarrow$/$\tau_\uparrow$$\ll$1), in exact similarity
to the power-dependent experiments at 10K.

The asymmetry of the capture and escape mechanisms is in fact the
fundamental reason why motional narrowing strikingly occurs when
decreasing the incident power or the temperature. If both
processes had the same efficiency
($\tau_\uparrow$$\sim$$\tau_\downarrow$), we would have
$\tau_c$$\sim$$\tau_\downarrow/2$ and $\Sigma$$\sim$$\Sigma_s$
which means that the spectral modulation amplitude would not
depend on the time constant $\tau_\downarrow$. Therefore, the
ratio $\Sigma\tau_c$/$\hbar$ could only \textit{decrease when
increasing the reservoir excitation}. This situation corresponds
to the well-known phenomenology from nuclear magnetic resonance
where the activation of the nuclei motion induces the motional
narrowing effect. In the present case where
$\tau_\downarrow$/$\tau_\uparrow$$\leq$10$^{-2}$, we are in the
opposite regime where the correlation time is merely constant with
relative variation smaller than 10$^{-2}$ whereas the spectral
modulation amplitude shows a steep increase with
$\tau_\downarrow$/$\tau_\uparrow$
($\Sigma$$\propto$$\sqrt{\tau_\downarrow/\tau_\uparrow}$). The
ratio $\Sigma\tau_c$/$\hbar$ thus \textit{increases when
increasing the reservoir excitation}.

We conclude that motional narrowing in the optical spectrum of a
quantum dot broadened by spectral diffusion occurs in the
unexpected regime of low temperature and low incident power. This
result is of great importance for the applications of quantum dots
to quantum information processing. In fact, our study demonstrates
that the decoherence dynamics can still be characterized by a
decoherence time $T_2$ even in the presence of fluctuating local
electric fields causing spectral diffusion. In that case, the
zero-phonon line has a Lorentzian profile and its width is not
given by the intrinsic radiative limit, but by the extrinsic
reservoir fluctuation dynamics. In particular, we provide a
straightforward interpretation to the widely debated issue of the
origin of the temperature dependence of the zero-phonon linewidth.
Eventually, we show that the motional narrowing effect strongly
reduces the environment-induced dephasing since the zero-phonon
linewidth is approximately $\Sigma\tau_c$/$\hbar$ smaller than the
spectral modulation amplitude $2\sqrt{2ln2}$\,$\Sigma$ due to
spectral diffusion. This unconventional motional narrowing opens a
novel route towards the control of environment-induced dephasing
for the applications of quantum dots to quantum information.

$^\ast$Electronic address: Guillaume.Cassabois@lpa.ens.fr

References

\vspace{1cm}
Acknowledgements

We thank C. Ciuti and B. Gayral for helpful discussions. LPA-ENS
is "unit\'{e} mixte (UMR 8551) de l'ENS, du CNRS, des
Universit\'{e}s Paris 6 et 7". This work is financially supported
by the region Ile de France through the project SESAME E-1751.

\vspace{1cm}
Competing financial interest

The authors declare that they have no competing financial
interests.


\begin{references}
\bibitem{bloembergen}Bloembergen, N., Purcell, E. M. \& Pound, R. V. Relaxation effects in nuclear magnetic resonance, Phys. Rev. \textbf{73}, 679-715 (1948).
\bibitem{kubo}Kubo, R. A stochastic theory of line-shape and relaxation, in \textit{Fluctuation, Relaxation and Resonance in Magnetic Systems} (ed D. Ter Haar) 23-68 (Oliver and Boyd, Edinburgh,
1962).
\bibitem{dyakonov}Dyakonov, M. I. \& Perel, V. I. Spin orientation of electrons associated with the interband absorption of light in semiconductors, Sov. Phys. JETP {\bf 33}, 1053-1059 (1971).
\bibitem{oxtoby}Oxtoby, D. W. Hydrodynamic theory of vibrational dephasing in liquids, J. Chem. Phys. {\bf 70}, 2605-2610 (1979).
\bibitem{eberly}Eberly, J. H., W\'{o}dkiewicz, K. \& Shore, B. W. Noise in strong laser-atom interactions: phase telegraph noise, Phys.
Rev. A \textbf{30}, 2381-2389 (1984).
\bibitem{flach}Flach, R., Hamilton, D. S., Selzer, P. M. \& Yen, W. M. Time-Resolved Fluorescence Line-Narrowing Studies in LaF$_3$:Pr$^{3+}$, Phys.
Rev. Lett. \textbf{35}, 1034-1037 (1975).
\bibitem{moerner}Ambrose, W. P. \& Moerner, W. E. Fluorescence spectroscopy and spectral diffusion of single impurity molecules in a crystal, Nature \textbf{349}, 225-227 (1991).
\bibitem{bawendi}Empedocles, S. A., Norris, D. J. \& Bawendi, M. G. Photoluminescence spectroscopy of single CdSe nanocrystallite quantum dots, Phys.
Rev. Lett. \textbf{77}, 3873-3876 (1996).
\bibitem{robinson}Robinson, H. D. \& Goldberg, B. B. Light-induced spectral diffusion in single self-assembled quantum dots, Phys.
Rev. B \textbf{61}, R5086-5089 (2000).
\bibitem{besombes}Besombes, L., Kheng, K., Marsal, L. \& Mariette, H. Few-particle effects in single CdTe quantum dots, Phys. Rev. B \textbf{65}, 121314-121317 (2002).
\bibitem{kammerer}Kammerer, C., Cassabois, G., Voisin, C., Perrin, M., Delalande, C., Roussignol, Ph. \& G\'{e}rard, J. M. Interferometric correlation spectroscopy in single quantum dots, Appl. Phys. Lett. {\bf 81}, 2737-2739 (2002).
\bibitem{shah}J. Shah, in \textit{Ultrafast spectroscopy of semiconductors and semiconductor nanostrutures} (Spinger-Verlag, Berlin, 1998).
\bibitem{gerard}G\'{e}rard, J. M., Genin, J. B., Lefebvre, J., Moison, J. M., Lebouche, N. \& Barthe, F. Optical investigation of the self-organized growth of InAs/GaAs quantum boxes, J. Cryst. Growth \textbf{150}, 351-356 (1995).
\bibitem{heitz}Heitz, R., Ramachandran, T. R., Kalburge, A., Xie, Q., Mukhametzhanov, I., Chen, P. \& Madhukar, A.
Observation of reentrant 2D to 3D morphology transition in highly
strained epitaxy: InAs on GaAs, Phys. Rev. Lett. \textbf{78},
4071-4074 (1997).
\bibitem{kaiser}Lohner, A., Woerner, M., Elsaesser, M. \& Kaiser, W. Picosecond capture of photoexcited holes by shallow acceptors in p-type GaAs, Phys. Rev. Lett. \textbf{68}, 3920-3923 (1992).
\bibitem{bimolecular}J. E. Carroll, in \textit{Rate equation in semiconductor electronics} (Cambridge University Press, 1985).
\bibitem{auger}O'Hara, K. E., Gullingsrud, J. R. \& Wolfe, J. P. Auger decay of excitons in Cu$_2$O, Phys. Rev. B {\bf 60}, 10872-10885 (1999).
\bibitem{besombes2}Besombes, L., Kheng, K., Marsal, L. \& Mariette, H. Acoustic phonon broadening mechanism in single quantum dot emission, Phys. Rev. B {\bf 63}, 155307-155311 (2001).
\bibitem{favero}Favero, I., Cassabois, G., Ferreira, R., Darson, D., Voisin, C., Tignon, J., Delalande, C., Bastard, G., Roussignol, Ph. \& G\'{e}rard, J. M. Acoustic phonon side-bands in the emission line of single InAs/GaAs quantum dots, Phys. Rev. B {\bf 68}, 233301-233304 (2003).
\bibitem{jusserand}Jusserand, B. \& Cardona, M., in \textit{Light scattering in solids V} (eds Cardona, M. \& Guntherrodt G.) 49-152 (Springer-Verlag, Heidelberg, 1989).
\end{references}
\end{document}